\documentclass[osa,preprint]{revtex4b5}

\input epsf
\begin{document}
\title{Wavy stripes and squares in zero Prandtl number convection}
\author{Pinaki Pal and Krishna Kumar}
\affiliation{Physics and Applied Mathematics Unit, Indian
Statistical Institute, 203, B.~T.~Road,  Calcutta-700~035, India}
\date{\today}
\begin{abstract}
A simple model to explain numerically observed behaviour of
chaotically varying stripes and square patterns in zero
Prandtl number convection in Boussinesq fluid is presented. 
The nonlinear interaction of mutually perpendicular sets of 
wavy rolls, via higher order modes, may lead to a 
competition  between the two sets of rolls. The appearance 
of square patterns is due to secondary forward Hopf 
bifurcation of a set of wavy rolls.
\pacs{PACS number(s):47.27.-i}
\end{abstract}
\maketitle

The study of low Prandtl number thermal convection~[1-13] has
been long motivated by its importance in geophysical and
astrophysical problems. The hydrodynamical equations of thermal
convection in Boussinesq fluids involve two types of
nonlinearity. The first  describes self-interaction ${\bf v .
\nabla v}$ of the velocity field ${\bf v}$, and the second
nonlinearity ${\bf v . \nabla} \theta $ results from the
advection of the temperature fluctuation $\theta$ by the
velocity field. The nonlinearity ${\bf v . \nabla} \theta $
may be neglected in the asymptotic limit of zero Prandtl
number~\cite{spiegel}. The linearly growing two-dimensional
(2D) rolls  then become exact solutions of nonlinear
equations, if {\it stress-free} boundary conditions are
considered. This makes this limit interesting even from purely
theoretical point. Thual~\cite{thual} recently showed by  a most
general 3D direct numerical simulations (DNS) of zero $P$
asymptotic equations that the solutions do not blow up.
The comparison with full Boussinesq equations with both
nonlinearities also reproduced zero P results. This DNS
showed many interesting patterns including the  possibility
of square patterns in zero $P$ Boussinesq fluids with
stress-free boundaries for $1.05 < r (=R/R_c) < 1.7$, where
the critical Rayleigh number $R_c = 27\pi^4/4 $. However,
the mechanism of generation of square patterns in zero $P$
convection remains unexplained. A nonlinear interaction
between 2D rolls cannot generate either square or hexagonal
patterns~\cite{kumar} for zero $P$ asymptotic equations as
growing straight 2D rolls are their exact solutions. The
streamlines in DNS~\cite{thual} support this view. On the
other hand, the mechanism of prevention of continuous growth
of  2D rolls approximately $0.5 \%$ above the onset of
convection was captured in a simple dynamical
system~\cite{kft}, which agreed well with the results of DNS
in its validity range. This model suggested that 2D rolls
undergo {\it self-tuned} nonlocal wavy instability~\cite{kft},
which prevents their further growth. The nonlinear
superposition of two sets of wavy rolls may result in the
form of square patterns. However, this proposition is not
analyzed so far.

We present in this article a simple dynamical system,
which describes nonlinear interaction between mutually
perpendicular sets of wavy rolls, and captures the
mechanism of selection of square patterns in zero Prandtl
number Boussinesq fluid. We show that the
generation of vertical vorticity is important in addition to
higher order modes to provide nonlinear coupling among
mutually perpendicular sets of  rolls. This is
qualitatively different from the case of high Prandtl number
convection~\cite{das} where nonlinear interaction of
two sets of straight rolls may give rise to square patterns
even in absence of vertical vorticity. The mutually
perpendicular sets of wavy rolls interact through distortions in
the vertical velocity as well as the vertical vorticity.
The nonlinear interaction  give rise to complex convective
patterns. The sequence of wavy stripe along $x-$axis, square
patterns, and $y-$ axis  is chaotic. The generation of
square patterns from wavy rolls is via forward Hopf
bifurcation.

We consider a thin horizontal layer of fluid of thickness $d$,
confined between two conducting boundaries, and heated
underneath. The fluid motion is assumed to be governed by zero
Prandtl number Boussinesq equations~\cite{{spiegel},{thual}},
which may be put in the following dimensionless form:
\begin{eqnarray}
\partial_t{\nabla^2v_3} &=& {\nabla}^4v_3 +R{{\nabla}_H}^2\theta \nonumber\\
 &-& {\bf{e_3}}.{\bf{\nabla}}\times[({\bf{\omega}}.{\bf {\nabla}}){\bf v} -
({\bf v}.{\bf{\nabla}}){\bf{\omega}}]\\
\partial_t{\omega_3} &=& {\nabla}^2{\omega_3} + [({\bf{\omega}}.{\bf
 {\nabla}}){v_3}-({\bf v}.{\bf{\nabla}}){\omega_3}]\\
 {\nabla}^2\theta &=& -{v_3}
\end{eqnarray}
where ${\bf v} \equiv (v_1,v_2,v_3)$ is the velocity field, $\theta$
the deviation from the conductive temperature profile, and
$\omega={\nabla}\times{\bf v}\equiv (\omega_1,\omega_2, \omega_3)$
the vorticity field. The Rayleigh number R is
defined as $R = \frac{\alpha(\Delta T)g{d}^3}{\nu\kappa}$, where $\alpha$ is the
coefficient of thermal expansion of the fluid, g the acceleration due to
gravity. The unit vector ${\bf e_3}$ is directed vertically upward, which
is assumed to be the positive direction of $x_3$-axis. The boundary
conditions at the stress-free conducting flat surfaces imply
$\theta = v_3 = \partial_{33}{v_3} = 0$ at $x_3 = 0, 1$.
and ${{\nabla}_H}^2 = \partial_{11} + \partial_{22}$ is the
horizontal Laplacian.

We employ the standard Galerkin procedure to describe the
convection patterns in the form two sets mutually perpendicular
sets of wavy rolls, and the patterns resulting due to their
nonlinear superposition. The spatial dependence of the vertical
velocity and the vertical vorticity are expanded in Fourier
series, which is compatible with the stress-free flat conducting
boundaries and periodic boundary conditions in the horizontal
plane.  As DNS shows either standing waves or stationary
patterns, all time-dependent Fourier amplitudes are set to be
real. The expansion for all the fields are truncated to describe
convective structures  in the form of straight cylindrical (2D)
rolls, wavy (3D) rolls, and patterns arising due to their
nonlinear superposition. The vertical velocity $v_3$ and the
vertical vorticity $\omega_3$ then read as
\begin{eqnarray}
v_3 &=& [W_{101}(t) \cos{(kx_1)} +
W_{011}(t)\cos{(kx_2)}]\sin{(\pi x_3)}\nonumber\\
&+& W_{211}(t) \cos{(2kx_1)}\cos{(kx_2)}\sin{(\pi
x_3)}\nonumber\\ &+&
W_{121}(t)\cos{(kx_1)} \cos{(2kx_2)}\sin{(\pi x_3)}\nonumber\\
&+& W_{111}(t) \sin{(kx_1)} \sin{(kx_2)} \sin{(\pi
x_3)}\nonumber\\
&+& W_{112}(t) \cos{(kx_1)} \cos{(kx_2)} \sin{(2\pi x_3)} +....\\
\omega_3 &=& Z_{010}(t) \cos{(kx_2)} + Z_{100}(t)
\cos{(kx_1)}\nonumber\\
&+& Z_{111}(t) \cos{(kx_1)} \cos{(kx_2)} \cos{(\pi x_3)}\nonumber\\
&+& [Z_{201}(t)\cos{(2kx_1)} + Z_{021}(t)\cos{(2kx_2)}] \cos{(\pi
x_3)}\nonumber\\
&+& [Z_{102}(t)\cos{(kx_1)}  + Z_{012}(t)\cos{(kx_2)}]\cos{(2\pi
x_3)}\nonumber\\
&+& Z_{210}(t)\cos{(2kx_1)}\cos{(kx_2)} \nonumber\\
&+& Z_{120}(t)\cos{(kx_1)}\cos{(2kx_2)}+....
\end{eqnarray}
The temperature field is slaved in the limit of zero Prandtl
number, and it can be computed from Eq. 3. The horizontal
components $v_1, v_2$ of the velocity  field and the same
($\omega_1, \omega_2$) of the vorticity field are then computed
by their solenoidal character. We now project the hydrodynamical
equations $1-3$ into these modes to get the following dynamical
system
\begin{eqnarray}
\dot{\vec{X}} &=& \frac{3\pi^2}{2}(r-1){\vec{X}}+\frac{1}{6}\left
(\begin{array}{c}{S\zeta_1}\\-{S\zeta_2}\end{array}\right)
+\frac{1}{4}\left(\begin{array}{c}{S\phi_1}\\-{S\phi_2}\end{array}
\right)+\frac{1}{20}\left(\begin{array}{c}{S\psi_1}\\-{S\psi_2}\end{array}\right)
\nonumber\\&-&\frac{1}{3\pi}{\eta}{\vec{\zeta}}-
\frac{1}{6\pi}{\eta}{\vec{\phi}}-\frac{1}{30\pi}
{\eta}{\vec{\psi}}-\frac{2}{15\pi}\left(\begin{array}{c}
{\chi_2}{\psi_2}\\{\chi_1}{\psi_1}\end{array}\right)
-\frac{\pi}{24}T{\vec{Y}}+\frac{\pi}{4}\left(\begin{array}{c}
{TX_2}\\{TX_1}\end{array}\right)\\
\dot{S} &=&\frac{\pi^2}{16}(27r-32)S-{\zeta_1}{X_1}+{\zeta_2}{X_2}
+\frac{1}{10}(-{X_1}{\psi_1}+{X_2}{\psi_2})+\frac{1}{2}
(-{X_1}{\phi_1}+{X_2}{\phi_2})\nonumber\\&+&\frac{3}{100}({\psi_1}{Y_2}
-{\psi_2}{Y_1})+\frac{3}{20}({\phi_1}{Y_2}-{\phi_2}{Y_1})
+\frac{3}{10}({\zeta_1}{Y_2}-{\zeta_2}{Y_1})\\
\dot{T} &=&\frac{\pi^2}{100}(27r-500)T - \frac{3\pi}{5}{X_1}{X_2} -
\frac{13\pi}{50}\vec{X}.\vec{Y}+\frac{3}{40}S({\chi_1}-{\chi_2})
-\frac{63\pi}{500}{Y_1}{Y_2}\nonumber\\&-&\frac{1}{10\pi}\eta({\chi_1}+{\chi_2})
-\frac{2}{25\pi}({\psi_1}{\phi_2}+{\psi_2}{\phi_1})
-\frac{4}{5\pi}({\zeta_2}{\phi_1}+{\zeta_1}{\phi_2})\\
\dot{\vec{Y}} &=&\frac{\pi^2}{98}(135r-343)\vec{Y}+\frac{31\pi}{28}
T\vec{X}+\frac{3}{28}\left(\begin{array}{c}{S\phi_2}\\-{S\phi_1}
\end{array}\right)+\frac{9\pi}{40}\left(\begin{array}{c}T{Y_2}\\T{Y_1}
\end{array}\right)\nonumber\\ &-&\frac{1}{14\pi}\left(\begin
{array}{c}{\eta}{\phi_2}\\{\eta}{\phi_1}\end{array}\right)+\frac{3}{28}
\left(\begin{array}{c}{S\psi_2}\\-{S\psi_1}\end{array}\right)
-\frac{9}{70\pi}\left(\begin{array}{c}{\eta}{\psi_2}\\{\eta}{\psi_1}
\end{array}\right)-\frac{8}{35\pi}\left(\begin{array}{c}{\chi_2}{\psi_1}
\\{\chi_1}{\psi_2}\end{array}\right)\nonumber\\
&+&\frac{9}{14}\left(\begin{array}{c}{S\zeta_2}\\-{S\zeta_1}\end{array}
\right)-\frac{1}{7\pi}\left(\begin{array}{c}{\eta}{\zeta_2}\\{\eta}{\zeta_1}\end{array}
\right)-\frac{2}{7\pi}\left(\begin{array}{c}{\chi_1}{\phi_1}\\
{\chi_2}{\phi_2}\end{array}\right)-\frac{4}{7\pi}
\left(\begin{array}{c}{\zeta_1}{\chi_1}\\{\zeta_2}{\chi_2}
\end{array}\right)\\
\dot{\vec{\zeta}} &=&-\frac{\pi^2}{2}\vec{\zeta}+\frac{\pi^2}{8}
\left(\begin{array}{c}SX_{1}\\-SX_{2} \end{array} \right)
+\frac{3\pi^2}{80}\left(\begin{array}{c}SY_{2}\\-SY_{1} \end{array} \right)
+\frac{\pi}{8}{\eta}{\vec{X}}-\frac{\pi}{8}
\left(\begin{array}{c}{\chi_2}{X_2}\\{\chi_1}{X_1}\end{array}\right)\nonumber\\
&-&\frac{\pi}{80}\left(\begin{array}{c}{\eta}{Y_2}\\{\eta}{Y_1}\end{array}\right)
+\frac{\pi}{4}\left(\begin{array}{c}T{\phi_2}\\T{\phi_1}\end{array}\right)
+\frac{\pi}{20}\left(\begin{array}{c}{\chi}_{1}Y_{1}\\{\chi}_{2}X_{2}
\end{array}\right)\\
\dot{\eta} &=& -2{\pi^2}{\eta}+\pi{\vec{\zeta}}.{\vec{X}}
+\frac{3\pi}{2}\vec{\phi}.\vec{X}
-\frac{\pi}{2}{\vec{\psi}}.{\vec{X}}
+\frac{\pi}{8}T({\chi_1}+{\chi_2})
-\frac{7\pi}{20}({\phi_2}{Y_1}+{\phi_1}{Y_2})\nonumber \\
&+&\frac{3\pi}{10}({\zeta_2}{Y_1}+{\zeta_1}{Y_2})
+\frac{\pi}{20}({\psi_1}{Y_2}+{\psi_2}{Y_1})\\
\dot{\vec{\psi}} &=& -\frac{5\pi^2}{2}{\vec{\psi}}
+\frac{\pi^2}{8}\left(\begin{array}{c}-SX_{1}\\SX_{2} \end{array} \right)
+\frac{7\pi}{8}{\eta}{\vec{X}}
+{\pi}\left(\begin{array}{c}{\chi_1}X_{2}\\{\chi_2}X_{1}\end{array}\right)
+\frac{9\pi_2}{80}\left(\begin{array}{c}S{Y_2}\\{-SY_1}\end{array}\right)\nonumber\\
&+&\frac{5\pi}{4}\left(\begin{array}{c}T{\phi_2}\\T{\phi_1}\end{array}\right)
+\frac{13\pi}{80}\left(\begin{array}{c}{\eta}{Y_2}\\{\eta}{Y_1}\end{array}\right)
+\frac{11\pi}{40}\left(\begin{array}{c}{\chi_2}{Y_1}\\{\chi_1}{Y_2}\end{array}\right)\\
\dot{\vec{\phi}} &=& -\frac{9\pi^2}{2}{\vec{\phi}}
+\frac{\pi^2}{8}\left(\begin{array}{c}{-SX_1}\\{SX_2}\end{array}\right)
-\frac{3\pi}{8}\left(\begin{array}{c}{\chi_2}{X_2}\\{\chi_1}{X_1}\end{array}\right)
-\frac{\pi}{8}\eta{\vec{X}}+\frac{\pi}{2}\left(\begin{array}{c}{T\zeta_2}
\\{T\zeta_1}\end{array}\right)\nonumber\\&-&\frac{\pi}{4}\left(\begin{array}{c}
T{\psi_2}\\T{\psi_1}\end{array}\right)+\frac{3\pi_2}{80}
\left(\begin{array}{c}{-SY_2}\\{SY_1}\end{array}\right)+\frac{9\pi}{80}
\left(\begin{array}{c}\eta{Y_2}\\\eta{Y_1}\end{array}\right)+\frac{\pi}{20}
\left(\begin{array}{c}{\chi_1}{Y_1}\\{\chi_2}{Y_2}\end{array}\right)\\
%\end{eqnarray}
%\end{document}
\dot{\vec{\chi}} &=& -3{\pi^2}{\vec{\chi}}+\frac{3\pi}{2}\left
(\begin{array}{c}{\phi_2}{X_1}\\{\phi_1}{X_2}\end{array}\right)
+\frac{3\pi^2}{8}\left(\begin{array}{c}{-ST}\\{ST}\end{array}\right)+
\frac{3\pi}{8}\left(\begin{array}{c}{\eta T}\\{\eta T}\end{array}\right)\nonumber\\&+&\pi
\left(\begin{array}{c}{\zeta_2}{X_1}\\{\zeta_1}{X_2}\end{array}\right)
+\frac{\pi}{5}\left(\begin{array}{c}{\phi_1}{Y_1}\\{\phi_2}
{Y_2}\end{array}\right)+\frac{2\pi}{5}\left(\begin{array}{c}{\zeta_1}{Y_1}\\
{\zeta_2}{Y_2}\end{array}\right)+\frac{\pi}{10}\left(\begin{array}{c}
{\psi_2}{Y_2}\\{\psi_1}{Y_1}\end{array}\right)
\end{eqnarray}
where
${\vec{X}}$ $\equiv$ $\left(\begin{array}{c} X_1\\X_2 \end{array} \right)$ $= $
$\left(\begin{array}{c} {W_{101}}\\{W_{011}} \end{array} \right)$,
${\vec{Y}}$ $\equiv$ $\left(\begin{array}{c} Y_1\\Y_2 \end{array} \right)$ $=$
$\left(\begin{array}{c} {W_{211}}\\{W_{121}} \end{array} \right)$,
$S=W_{111}$, $T=W_{112}$ ,
${\vec{\zeta}}$ $\equiv$ $\left(\begin{array}{c}{\zeta_1}\\{\zeta_2}\end{array} \right)$ $=$
$\left(\begin{array}{c} {Z_{010}}\\{Z_{100}} \end{array} \right)$,
${\vec{\phi}}\equiv \left(\begin{array}{c}{\phi_1}\\{\phi_2}\end{array} \right)$ $=$
$\left(\begin{array}{c} {Z_{012}}\\{Z_{102}} \end{array} \right)$,
${\vec{\psi}}$ $\equiv$ $ \left(\begin{array}{c}{\psi_1}\\{\psi_2}\end{array} \right)$ $=$
$\left(\begin{array}{c} {Z_{210}}\\{Z_{120}} \end{array} \right)$,
${\vec{\chi}}$ $\equiv$ $ \left(\begin{array}{c}{\chi_1}\\{\chi_2}\end{array} \right)$ $=$
$\left(\begin{array}{c} {Z_{201}}\\{Z_{021}} \end{array} \right),$
and ${\eta}={Z_{111}}$. All roman characters stand for
velocity modes, and greek letters stand for vorticity modes.
\begin{figure}[t]
\hspace*{2.0cm}
\centerline{\epsfxsize=3.0in\epsfbox{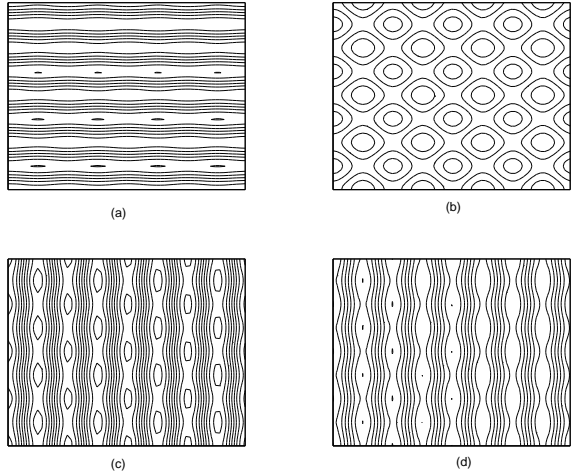}}
\caption{Four different patterns varying chaotically in time for $r=1.11$:
(a) wavy rolls along x-axis, (b) square pattern, (c) patchwork
quilt pattern , and (d) wavy rolls along y-axis.}
\label{fig1}
\end{figure}

\begin{figure}
\hspace*{2.0cm}
\centerline{\epsfxsize=3.0in\epsfbox{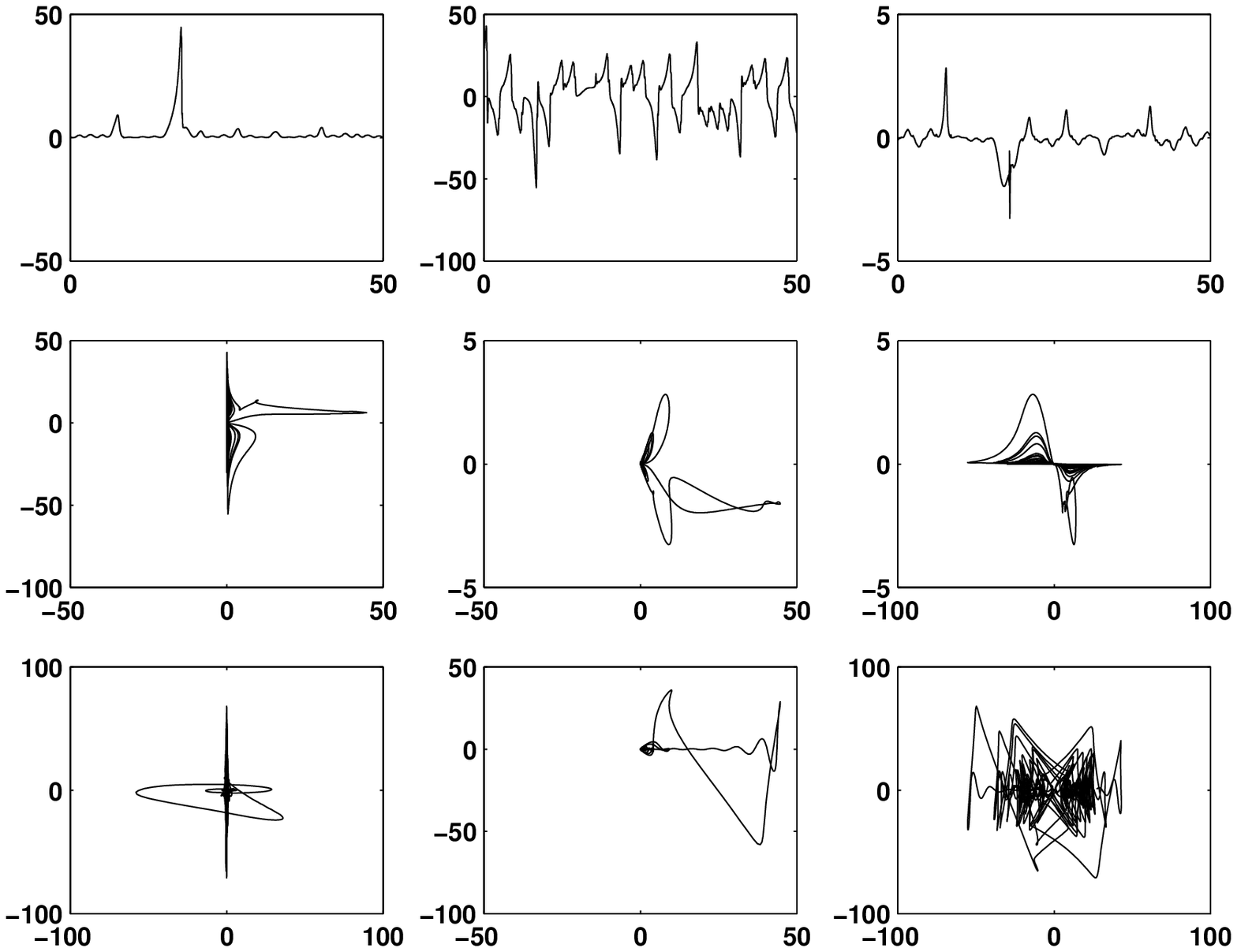}}
\caption{The top row (starting from left) shows chaotic evolution of $W_{101}$,
$W_{011}$, and $W_{112}$ for $r=1.11$. The second row shows
projections of phase space plot in  $W_{011} - W_{101}$,
$W_{112} - W_{101}$, and $W_{112} - W_{011}$ planes respectively.
The third row contains projections of phase space plot
in $Z_{100} - Z_{010}$, $Z_{010} - W_{101}$, and
$W_{011} - Z_{100}$ planes respectively.}
\label{fig2}
\end{figure}

There is no fixed points in this model as growing straight rolls
are exact solutions of the problem. We now numerically integrate
this dynamical system to investigate time dependent solutions.
For each value of $r$, we start integration with  randomly chosen
initial conditions. We integrate for long periods ($\approx 100 -
200 $ times the viscous diffusive time scale) to ignore the
transient effects. We increase $r$ in small steps $\Delta r$
($=0.01$) and repeat the above mentioned procedure.  We get
chaotic solutions for $1.11 \le r \le 1.42$ for $k_x = k_y = k_c
= \pi/\sqrt{2}$, which is  in qualitative agreement with the DNS
of Thual~\cite{thual}. For $r < 1.11$ we observe wavy rolls
oscillating chaotically, and for $r \ge 1.11$ we observe chaotic
sequence of wavy rolls, squares, and  asymmetric squares. For $r
> 1.42$ and $k_x = k_y = k_c$, the model is not good enough as
the anharmonicity in various fields is very high in the limit of
zero Prandtl number. This is natural as there is no additional
heat flux across the fluid layer due to convective motion. The
extra energy is consumed by the internal dynamics.

Figure 1 shows competition of various patterns for $r=1.11$. Two
mutually perpendicular sets of wavy rolls compete with each
other. This leads to  square  as well as patchwork quilt
patterns. The sequence of occurrence of these patterns is {\it
not} periodic but chaotic. Notice that the square pattern (see
Fig. 1) consists of two sets of squares. A small (big) square has
four big (small) squares as its nearest neighbors (Fig. 1 b).
This feature of the square pattern is qualitatively new for
convection in Boussinesq fluid.

The top row of Fig. 2 (starting from left) shows time
evolution of velocity modes $W_{101}$, $W_{011}$ and
$W_{112}$ respectively after integration of the model
for a period more than $200 \times$ viscous time scale.
The signal is chaotic. The second and third rows of
Fig. 2 show various projections of the phase plots
in fifteen dimensional phase space. Figure 3 shows
time variation of the spatially average energy
$E_{av} = \frac{1}{2}<v_1^2 + v_2^2 + v_3^2>_{x_1 x_2 x_3}.$
The averaged energy shows relaxation oscillation in irregular
way. This is also a common feature of DNS~\cite{thual}.
\begin{figure}
\hspace*{2.0cm}
\centerline{\epsfxsize=2.5in\epsfbox{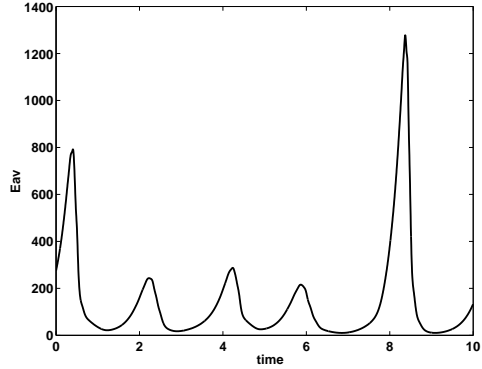}}
\caption{Time variation of spatially averaged energy in a unit
cell.}
\label{fig3}
\end{figure}

We have presented in this paper a simple model which explains the
mechanism responsible for generation of square patterns in zero
Prandtl number limit in Boussinesq fluids. As the Rayleigh number
is increased, one set of wavy rolls oscillating chaotically
become unstable. Another set of wavy rolls are generated in a
direction perpendicular to the former one. The nonlinear
superposition of these wavy rolls give rise to squares and other
complex patterns. The vertical vorticity modes with nonzero mean
in vertical direction are  responsible for the wavy nature of
rolls at very close to the instability onset in case of {\it
stress-free} bounding surfaces. These modes stop the unlimited
growth of the $2D$ rolls. The DNS~\cite{thual} also showed wavy
rolls rather than straight rolls. The nonlinear modes  depending
on both the horizontal coordinates  facilitate the exchange of
energy between two sets of wavy rolls.

%\newpage
\noindent
Acknowledgements: Pinaki Pal acknowledges support from CSIR, India
through its grants.

%\newpage
%\BibTex

%\begin{center}
% FIGURE CAPTION
%\end{center}

%\noindent
%Figure 1. Four different patterns varying chaotically in time for $r=1.11$:
%(a) wavy rolls along x-axis, (b) square patterns, (c) asymmetric square , and
%(d) wavy rolls along y-axis.\\

%\noindent
%Figure 2. The top row (starting from left) shows chaotic evolution of $W_{101}$,
%$W_{011}$, and $W_{112}$ for $r=1.11$. The second row shows
%projections of phase space plot in  $W_{011} - W_{101}$,
%$W_{112} - W_{101}$, and $W_{112} - W_{011}$ planes respectively.
%The third row contains projections of phase space plot
%in $Z_{100} - Z_{010}$, $Z_{010} - W_{101}$, and
%$W_{011} - Z_{100}$ planes respectively.\\

%\noindent
%Figure 3. Time dependence of the energy $<v_1^2 + v_2^2 + v_3^2>_{xyz}$.

\end{document}